# Statistical study of auroral omega bands

Noora Partamies[1,4], James M. Weygand[2], and Liisa Juusola[3]

[1]Department of Arctic Geophysics, University Centre in Svalbard, Longyearbyen, Norway
[2]University of Los Angeles, Department of Earth, Planetary and Space Sciences, California, USA
[3]Finnish Meteorological Institute, Earth Observations, Helsinki, Finland
[4]Birkeland Centre for Space Science, Norway

*Correspondence to:* Noora Partamies (noora.partamies@unis.no)

**Abstract.** The presence of very few statistical studies on auroral omega bands motivated us to test-use a semi-automatic method for identifying large-scale undulations of the diffuse aurora boundary and to investigate their occurrence. Five identical all-sky cameras with overlapping fields of view provided data for 438 auroral omega-like structures over Fennoscandian Lapland from 1996 to 2007. The results from this set of omega band events agree remarkably well with previous observations of omega band occurrence in magnetic local time (MLT), lifetime, location between the region 1 and 2 field-aligned currents, as well as current density estimates. The average peak emission height of omega forms corresponds to the estimated precipitation energies of a few keV, which experienced no significant change during the events. Analysis of both local and global magnetic indices demonstrates that omega bands are observed during substorm expansion and recovery phases that are more intense than average substorm expansion and recovery phases in the same region. The omega occurrence with respect to the substorm expansion and recovery phases is in a very good agreement with an earlier observed distribution of fast earthward flows in the plasma sheet during expansion and recovery phases. These findings support the theory that omegas are produced by fast earthward flows and auroral streamers, despite the rarity of good conjugate observations.

## 1 Introduction

Akasofu and Kimball (1964) were the first to describe the auroral wave-like structures called "omega bands", which appear within the morning sector auroral oval with shapes resembling the Greek letter Ω, and are typically associated with the recovery phase of magnetic substorms (e.g. Vanhamäki et al., 2009). However, a study by Wild et al. (2000) suggested that omega bands can be initiated at the substorm onset site and propagate eastward from there. To date, the generation mechanism for this auroral phenomenon has not been fully established. There are potentially three generation mechanisms in the magnetosphere: (1) omega bands form as a consequence of auroral streamer activity (Henderson et al., 2002, 2009, 2012; Henderson, 2016), where auroral streamers are ionospheric projections of earthward flow bursts in the plasma sheet (e.g. Nakamura et al., 2001) and auroral omega bands evolve from north–south-aligned streamers; (2) omega bands may arise through the structuring of magnetic vorticity and field-aligned currents in the ionosphere by the Kelvin–Helmholtz instability driven by flow shears at the inner edge of the plasma sheet (Rostoker and Samson, 1984; Janhunen and Huuskonen, 1993); (3) another magnetotail mechanism is that of Yamamoto et al. (1993, 1997) where omega bands are the result of perturbation of hot plasma torus boundary. Yamamoto et al. (1997) stated that the hot plasma torus system is potentially unstable: under certain conditions the electrostatic interchange instability due to the particle magnetic drifts can develop at the hot plasma torus boundary. In the latter two mechanisms omega bands form at the boundary of region 1 and 2 currents in the auroral oval, while in the first mechanism the streamer flow starts up within the region 1 current prior to the omega formation at the boundary of region 1 and 2 currents.

The largest statistical study of omega bands included about 600 automatically classified omega structures (Syrjäsuo and Donovan, 2004). The total of 350 000 Canadian all-sky camera images from Gillam station (at 56.37° geographic and 64.54° geomagnetic latitude) taken in 1993–1998 was analysed. The study included five mutually exclusive classes of (1) no aurora, (2) arcs, (3) patchy aurora (irregularly shaped emission patches), (4) omega bands ($\Omega$-shaped structures) and (5) other (e.g. diffuse or complex auroral structures). Excluding the first class left the shape analysis with 220 000 images containing aurora. The automatic classification was base d on brightness, alignment and multi-scale texture related features of auroral images. The training set included some tens (omegas) to more than 100 (arcs and patches) carefully selected samples per structure class. The automatic classification detected about 17 000 arcs, 9700 patchy auroras and 600 omega bands. Auroral structures were examined in single images without assessing their temporal evolution and lifetime. The classification of arcs and patches was concluded to be reliable, while the automatic detection of omega bands was found challenging. The number of omega bands maximize at about 02:30 MLT. Other properties of the detected omega structures were not examined in this classification-focused work.

A connection between the occurrence of omega bands and magnetic Ps6 pulsations has been shown by many studies (e.g. Wild et al., 2011, and references therein). Connors et al. (2003) studied magnetic Ps6 pulsations in the relation with substorm expansion onsets at midnight sector. Magnetic Ps6 pulsations (5–40 min) with amplitudes of about 10–1000 nT have been traditionally related to substorm recovery and auroral omega activity in the morning sector but Connors et al. (2003) showed that it is not uncommon to observe Ps6 signatures during the substorm expansion phase.

In a study of Amm et al. (2005), horizontal equivalent currents of the order of $1-2\,\mathrm{A\,m^{-1}}$, and field-aligned currents (FACs) of about $10-20\,\mathrm{A\,km^{-2}}$ were reported within a region of an auroral omega form. These rather extreme values were related to the analysed case during a geomagnetic storm with AE index of about $-1000$ nT and Dst index about $-100$ nT. The wavy structure in the horizontal equivalent currents moved together with the auroral omega band with an eastward propagation speed of several hundreds of m s$^{-1}$ in agreement with the $\boldsymbol{E}\times\boldsymbol{B}$ drift velocity. The westward flank of the omega form was observed to coincide with more intense auroral emission than the eastward one. Based on Polar UVI images they estimated average precipitation energies of 2–5 keV.

Most recently Weygand et al. (2015) carried on a detailed investigation of five omega band intervals with the total of 26 omega structures over the Time History of Events and Macroscale Interactions during Substorms (THEMIS) ground-based instrument network in Canada and Alaska. The lifetime of their omega bands ranged from 1.5 to 17 min. The structures occurred close to the boundary between region 1 and 2 currents in the post-midnight sector and some were observed to develop from auroral streamers. High-speed plasma sheet flows were measured by the THEMIS spacecraft prior to the ionospheric omega observations. Dipolarizations in the Geostationary Operational Environmental Satellite (GOES) magnetic field data at the same local time were also recorded. The plasma sheet high-speed flows were concluded as the most likely generation mechanism for the observed omega bands. The conclusion was supported by optical observations of the development of five omegas after auroral streamers. However, some omega structures evolved without any association with auroral streamers.

In this study, we show the typical behaviour of omega bands in terms of geomagnetic activity, structural evolution of the aurora, lifetime and occurrence as observed in the ground-based camera and magnetometer data. Space-borne observations of omega bands are used to describe the M–I coupling of the diffuse aurora boundary undulations whenever available.

## 2　Ground-based observations and event selection

The Magnetometers – Ionospheric Radars – All-sky Cameras Large Experiment (MIRACLE) network included five identical auroral all-sky camera (ASC) setups in the Fennoscandian Lapland in the period 1996–2007 (Sangalli et al., 2011). These imagers took pictures through fish-eye optics and optical filters for auroral green line (557.7 nm) every 20 s, as well as auroral blue (427.8 nm) and red (630.0 nm) line images every minute. Imaging required that the Sun was more than 10° below the horizon, which occurs during several hours every night from about September until about April. One winter season results in about 0.8 million images per year per station. The Lapland stations of Sodankylä (SOD, 67.42° N), Muonio (MUO, 68.02° N), Abisko (ABK, 68.36° N), Kilpisjärvi (KIL, 69.02° N) and Kevo (KEV, 69.76° N) have overlapping fields of view (FoV) which allow for correlation studies of the auroral structures between the neighbouring stations.

All MIRACLE ASC data from the era of 1996–2007 have been automatically pruned into classes of Aurora and No Aurora. The detection of the presence of aurora in this pruning procedure is based on thresholding: if the number of pixels above a local brightness threshold is sufficiently large, the image is assigned to contain aurora, as described in Syrjäsuo et al. (2001). This allows more efficient further analyses and searches in the image data. The pruning method with experimental parameters has been visually validated to detect practically all aurora seen in the keogram summary plots. A recently developed automatic analysis method by Whiter et al. (2013) triangulates auroral peak emission heights for all pruned data. As a side product of this method an auroral structure index "arciness" (Partamies et al., 2014) is calcu-lated. The arciness index describes the complexity of the auroral structures in an image based on clustering of the brightness distribution within the FoV. An auroral arc or a multiple arc in an image corresponds to arciness value of 1, while more complex structures result in lower arciness values down to about 0.4. A single value of peak emission height and arciness is assigned for each analysed image.

In order to find omega band events for the current analysis we first visually browsed randomly displayed pruned image data at one minute resolution from the five camera stations. This proved to be a fast way of detecting good samples of omega-shaped (or any type of) aurora. Secondly, a random projection method (Bingham and Mannila, 2001) was used to find similar structures in the image data. We used random projection of $32 \times 32$-pixel thumbnail images (1024 dimensions) projected onto 15 and 25 dimensions to capture the relevant information. An Eulerian distance in the numeric feature space was applied to find the images closest to our sample images (most similar). Thirdly, keograms and thumbnail images (http://www.gaia-vxo.org) were visually inspected for boundary undulations and quality of omega forms for the identified events which often resulted in new events being detected close to the inspected time range. A set of different colour scales were used to aid detection of the faintest features in the data. None of these search methods is ideal, but we believe that the combination of several searches has brought up the majority of the omega-like structures. This list of events can become valuable for developing computer vision methods in the future.

Properties required for omega band selection were (1) each omega must appear in more than one image in 1 min resolution thumbnail data, (2) each omega must look like the Greek $\Omega$ letter in at least one image (peak time), (3) each omega has to propagate east, (4) each omega must appear taller than wider at its peak time, and (5) each omega must fully fit in the camera FoV at least at one time point for reliable detection. All the omega-like structures have further been visually followed in the image data to find their start, end and peak times. The peak time refers to the time when the most Greek $\Omega$ letter-like structure has been observed well within the camera FoV.

With the procedure described above we found 438 omega-like structures fulfilling the requirements. Among those, there were 259 clear and distinct omega forms which are used as a reference group in our analysis. Another 179 omega forms were found to marginally fulfil the above requirements. They are typically slightly too faint to convincingly bring up the wavy structure (breaking criterion 2), too small (breaking criterion 4), too large (breaking criterion 5), too tall (breaking criterion 2), stationary (breaking criterion 3), only visible in one single image (breaking criterion 1), or partially visible (breaking criterion 2 and 5). Examples of distinct and less obvious omega forms can be seen in the middle and bottom panel of Fig. 1, respectively. The statistical results shown in this study are based on the analysis of all 438 omega-like structures after treating the whole event set and the reference set separately and finding them similar. Each omega form is treated as an individual in this statistical analysis, since the majority of the events (335) consisted of single omega forms, or omegas with more than half an hour time difference to the next one. Most omega-like structures were found in the data of the southernmost auroral camera, SOD. Table 1 lists the number of detected omega bands at each station together with the number of pruned images per station and the station coordinates.

An example omega band event in ASC data is shown in Fig. 1. These data are from SOD station on 2 September 2005. Three well-defined omega structures took place at 23:00–01:00 UT (bottom panel). These are seen as wave-like features of the diffuse aurora boundary in the keogram (marked with red vertical lines in the top panel). In addition to the marked structures (at 23:27, 00:27 and 00:50 UT, images in the middle panel), another three undulations were selected as omega-like structures (at 23:15, 00:39 and 00:47 UT, images in the bottom panel). Yet at least another three similar wave-like perturbations can be found in the same time frame without them fulfilling the required omega band properties.

To describe the ground-magnetic activity we use the Dst index from Kyoto World Data Center, local auroral electrojet index constructed from magnetograms at the five camera stations ($IL_{ASC}$; Partamies et al., 2015), and equivalent current calculated from all the magnetometer data in the IMAGE chain of magnetometers (35 stations at 10 s resolution in the period 1994–2014; Juusola et al., 2016).

## 3 Appearance and relation to magnetic activity

Most omega-like structures were found in the image data of the southernmost auroral camera, SOD (137 out of 438). This suggests an enhanced level of magnetic activity to widen the auroral oval to reach over the SOD station. The lifetimes of omega bands range from 1 to 47 min with a median value of 8 min (mean of 10 min). About 90 % of the omega forms are observed for less than 20 min. We consider the times to be underestimates of the true lifetimes since only their growth, drift or decay has been observed within the common FoV of the MIRACLE cameras. More than half of the omega-like forms occur at 00:00–02:00 UT, which is equivalent to 02:00–04:00 MLT (Fig. 2). This observation agrees very well with the peak occurrence at 02:30 MLT by Syrjäsuo and Donovan (2004). The sharp cutoff at around 04:00 MLT in Fig. 2 is most likely due to the typical end time of the auroral imaging in the Fennoscandian sector.

In addition to the occurrence versus MLT plot shown in Fig. 2 we examined the annual rate of omega band events. We found that the peak occurrence of omega bands took place in 2002, 2003 and 2004, in that order. Altogether, about half of all omega structures occurred during the years 2002–2004 in the declining phase of the solar cycle 23, which were also characterized by strong solar wind driving and high level of geomagnetic activity (e.g. Myllys et al., 2015).

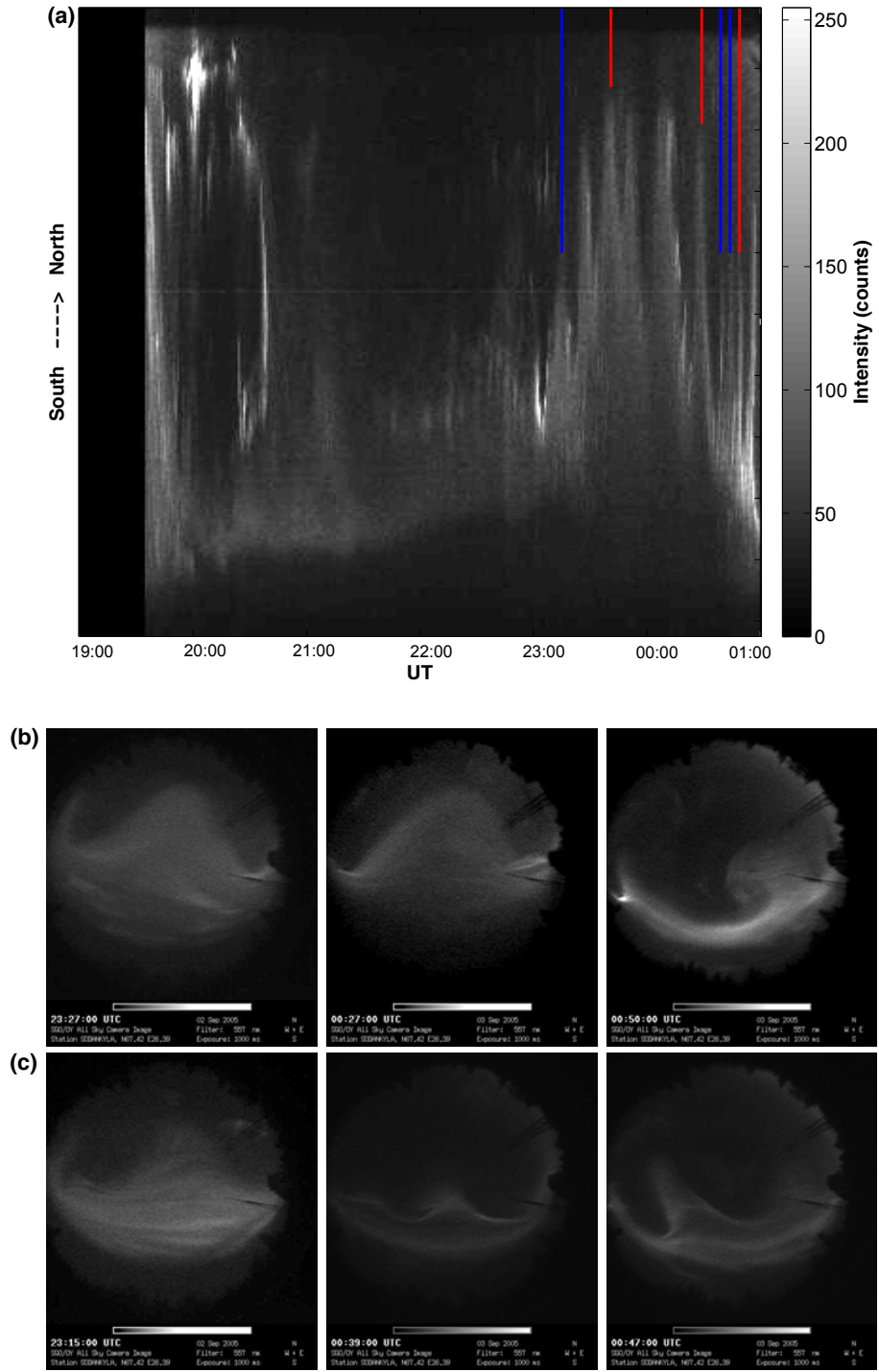

**Figure 1.** Keogram from SOD station on the night of 2 September in 2005 **(a)**. The $y$ axis is the zenith angle of the images from south (bottom) to north (top) and the $x$ axis is UT in hours. The intensity is in relative brightness units (counts from 0 to 255). At 23:00–01:00 UT the auroral boundary undergoes undulations which show up as wave-like features in the keogram. Three selected omega structures are shown in the middle panel images **(b)** taken at 23:27, 00:27 and 00:50 UT. The red vertical lines on the keogram mark the individual omega forms in **(b)**. Examples of less-obvious omega structures at 23:15, 00:39 and 00:47 UT are shown in **(c)** and marked by blue vertical lines in the keogram. The contrast in all images is enhanced.

Arciness is a number assigned for each auroral image and it indicates whether the auroral structure in the image is arc-like. A value of 1 for the arciness corresponds to an auroral arc or multiple arcs while a lower value indicates that the auroral structures are more complex with the brightness distribution more widespread throughout the image. The method is based on clustering of the brightest pixels in each image (Partamies et al., 2014). The median arciness decreases during the evolution of omegas from about 0.87 at 7 min before to about 0.81 at around the omega peak time (blue line in the left panel of Fig. 3). The normal range of arciness is 0.4–1. The observed change in median arciness during omega evolutions is small while the variations in arciness in general (as described by the black line quartiles in Fig. 3) are large. The key point to note, however, is that the higher arciness values prior to the omega structures generally do not recover within the 20 min time span after the omega peak time, and that the decreasing trend can be seen in the median as well as in the quartile values. The structural evolution after an omega has reached its most well-defined form tends to be more complex than that prior to the peak time. This suggests that the omega-like boundary undulation is not just a transient feature with a limited lifetime but it relates to a transition between relatively simple and more dynamic auroral displays.

Table 1. Names, geographic coordinates, corrected geomagnetic latitudes, years of operation and the total number of pruned images for the Lapland ASC stations (from south to north) used in this study. The magnetic midnight in Fennoscandia meridian is at about 21:30 UT.

| Station | Abbreviation | Glat | Glong | CGMlat | Years of operation | Pruned images | Omegas (distinct/all) |
|---|---|---|---|---|---|---|---|
| Sodankylä | SOD | 67.42° | 26.39° | 63.92° | 2000–2007 | 623 539 | 91/137 |
| Muonio | MUO | 68.02° | 23.53° | 64.72° | 1996–2007 | 450 099 | 31/66 |
| Abisko | ABK | 68.36° | 18.82° | 65.30° | 1997–2002 | 305 279 | 36/65 |
| Kilpisjärvi | KIL | 69.02° | 20.87° | 65.88° | 1996–2007 | 856 279 | 65/107 |
| Kevo | KEV | 69.76° | 27.01° | 66.32° | 1997–2006 | 555 303 | 36/83 |

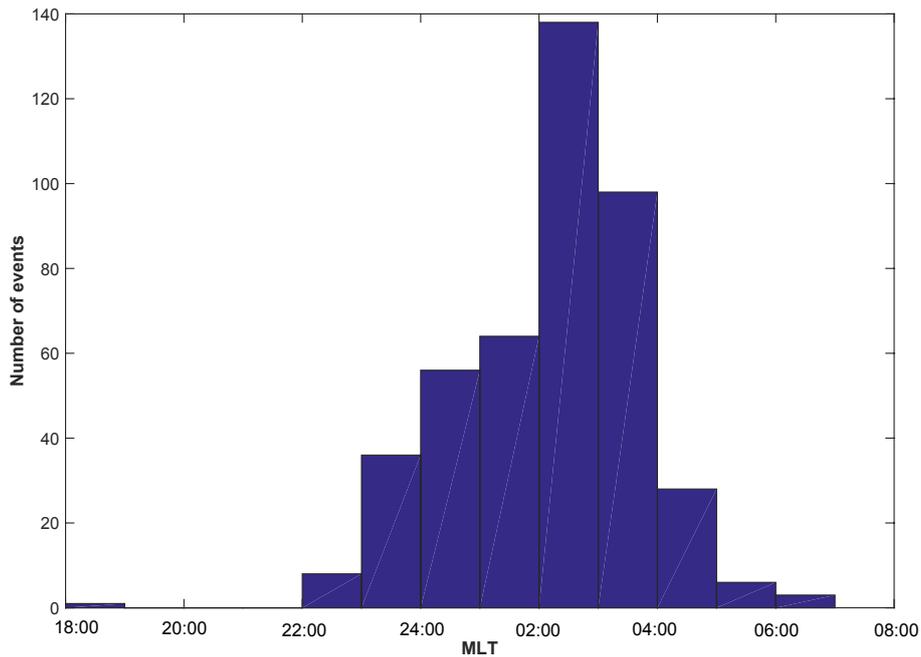

Figure 2. The magnetic local time distribution of omega-like structures has a peak value at 02:00–03:00 MLT.

The right panel of Fig. 3 shows the arciness evolution for substorm recovery phases as reference (data from Partamies et al., 2015). The zero epoch has been chosen for the beginning of the recovery phases, as that is when most of the omega structures have been observed. The median recovery phase evolution of the auroral structures shows a smooth variation of arciness values within the same range as those during the omegas (blue curve in the left panel). The difference is subtle and mainly relates to slightly more arc-like structures occurring prior to the omega passage (left) than at the same time in an average expansion phase (right), which may reflect the existence of a band of diffuse aurora prior to the omega formation. We also see a slower recovery of the arciness after the omega passage (left) than during an average recovery phase (right).

The median auroral peak emission height for omega-like structures is 118 km, which typically corresponds to precipitation energies of a few keV (Turunen et al., 2009) and is in a very good agreement with the previous energy estimates of 2–5 keV by Amm et al. (2005). The heights are estimated using an automatic triangulation-like method by Whiter et al. (2013). The median height (data not shown) shows no significant change over the lifetime of the omega structures suggesting that there is no major change in precipitation energy related to the passage of the omega band.

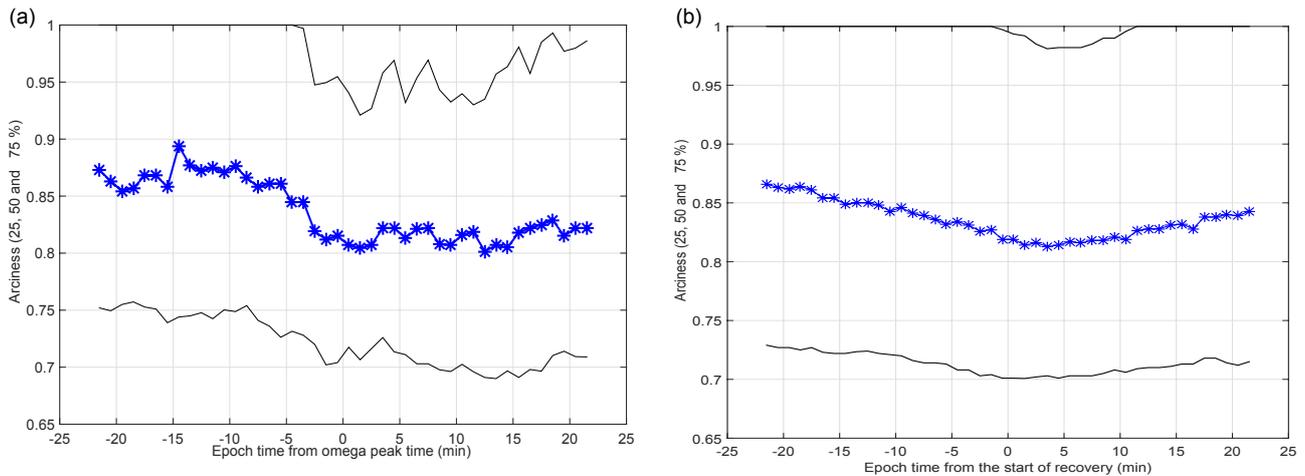

**Figure 3.** Median value of auroral arciness (blue curve) for ±20 min of the omega peak time for all 438 omega structures **(a)**, and for recovery phases **(b)**. The $x$ axis is the epoch time in minutes and the zero epoch time is the peak time of omegas, and the start of the phase for the recovery phases. The number of data points (images) in each 1 min time slots varies from 400 to 700 for omegas and is thousands for the recovery phases. Each omega form has been considered individually in the superposed epoch analysis. A high level of variation in the arciness is described by the quartiles (black curves) but the same mild behaviour is seen in the quartiles as well as in the medians.

Out of the reference set of 259 omega bands, two took place during growth phases, 96 during expansion phases and 161 during recovery phases. Taking all the omega-like structures the numbers for expansion and recovery phase structures become 165 and 274, respectively. The phases are automatically detected from the camera station IL index ($IL_{ASC}$) and IMF $B_Z$ by a search routine developed by Juusola et al. (2011) and applied to local electrojet index data by Partamies et al. (2015). The substorm expansion phase is defined as an abrupt decrease of the index value at a minimum rate of $4\,nT\,min^{-1}$. The recovery phases are defined to start from the end of the expansion phases and continue until the index value becomes higher than the long-term median value of $-50\,nT$. The local IL index has been shown to correspond well to, or perform better than, the global index in the nightside at about 22:30–04:00 MLT (Kauristie et al., 1996). We use a subset of stations included in IL to capture the magnetic variations within the average auroral oval latitudes and to look at the disturbances directly related to the omega bands. The time differences between the omega peak times and the substorm onsets (expansion phase starts) show an exponentially decaying distribution. In total, 90 % of all observed omega structures took place within 1.5 h from the substorm onset. That places most of the omega undulations into the recovery phase, since the mean duration for sub-storm expansion phases according to the local electrojet index data is of the order of 20 min (Partamies et al., 2015). The omega structures occurring during the expansion and recovery phases took place during phases which were longer than average. The two omega forms, which were observed during substorm growth phases, were found during short growth phases that were preceded by substorm activity rather than quiet time. Thus, the magnetospheric conditions for the two growth phase omega bands can be associated with substorm activity even though the next loading period was already in progress. These findings suggest that a substorm onset (tail reconnection) may be important in pre-conditioning the magnetosphere for the instability causing omega undulations.

In order to relate the omega structures to a certain fraction of a substorm phase, we normalized the durations of expansion and recovery phases to the range of 0–1. Figure 4 shows the distribution of peak times for all omega-like structures within the relative duration of the expansion (left panel) and recovery (right panel) phases. The variation is large but there is a tendency of omega-like forms to appear towards the end of the expansion (median at 57 % of expansion phase duration) and beginning of the recovery (median at 42 % of the recovery phase duration).

Median arciness values for substorm growth, expansion and recovery phases are 0.98, 0.84 and 0.82, respectively (Partamies et al., 2015). During the substorm phases where the omega structures have been observed the expansion phase arciness before and after the omega passage is 0.88 and 0.79, respectively. Similarly, recovery phase arciness before and after the omega passage is 0.85 and 0.84, respectively. Omega occurrence does not change the average structuring of the expansion phase aurora, but divides it into more arc-like and more complex. The omega passage does not change the morphology of the recovery phase aurora but results in a slightly more arc-like aurora than the average.

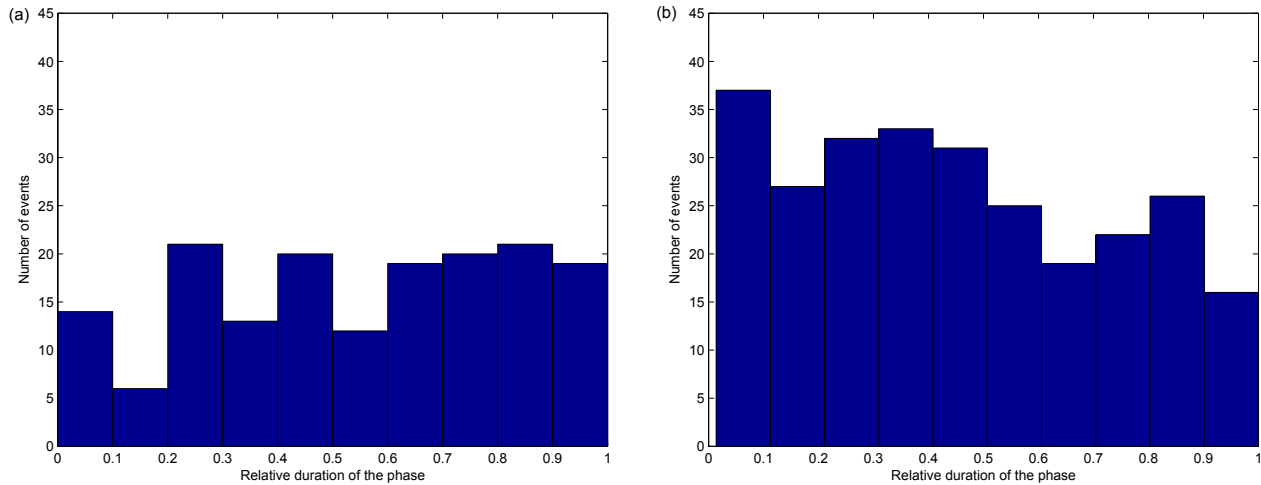

**Figure 4.** Occurrence of omega-like structures with respect to the relative duration of the substorm expansion **(a)** and recovery **(b)** phases. The substorm phases are automatically detected in the local electrojet index data $IL_{ASC}$. The total of 165 omegas was found within expansion and 274 of them within the recovery phases.

There were no significant systematic changes in global auroral electrojet index (AE/AL/AU) values during the omega evolution. The local electrojet index constructed from the five ASC station magnetograms ($IL_{ASC}$) shows the deflection related to the current enhancement during the passage of the omegas (left panel of Fig. 5), which equals about 40 nT over $\pm 20$ min around the omega peak time. Both the decrease prior to and the recovery after the omega peak time are slow and steady.

The observed median IL index values of about $-300$ to $-250$ nT are more than 100 nT lower than median values for all substorm expansion and recovery phases detected in the Lapland region ($\sim -160$ nT) (Partamies et al., 2015). Thus, the activity level at which omega forms are seen is more intense than that of typical substorm activity in this region. The local electrojet index shows no periodic decrease related to the omega events with several individual forms.

The Dst index is typically higher than $-50$ nT during the omega band events. The superposed epoch analysis in the right panel of Fig. 5 shows that the Dst index reaches a minimum about 5 h after the omega band events accompanied by a decrease of about 15 nT. The Dst values observed during omega bands are about 5–7 nT lower than those typically observed during auroral substorms in the Lapland region (Partamies et al., 2015). Furthermore, average Kp index value during omega bands is 4, which corresponds to sawtooth-type activity rather than the disturbance level associated with steady magnetospheric convections or isolated substorms (Partamies et al., 2009).

## 4 Equivalent currents

Equivalent current distributions from IMAGE magnetometer chain (Juusola et al., 2016) have been investigated for the reference set of distinct omega bands. An example equivalent current map during an omega band event is shown in Fig. 6.

The vector field in the maps illustrates the strength and direction of the equivalent current; the colour coding in the left panel gives the distribution of the curl of the equivalent currents, and the false-colour image in the right panel places the optically observed omega structure in the context of large-scale currents.

During the omega occurrences the strong westward electrojet (WEJ) region typically extends far into the evening sector and the omega bands take place within the southern part of it. This agrees with the fact that most omegas are observed at the southern part of the oval (SOD). The strength of the maximum horizontal current is over 500 A km$^{-1}$ for the examined omega cases. A long-term average of morning sector equivalent current is about 200 A km$^{-1}$ and an average maximum current during substorm expansions is around 400 A km$^{-1}$ (Juusola et al., 2015). The currents during omega events are thus clearly stronger than average equivalent current activity within the Fennoscandian sector. These previously published equivalent current numbers may be slightly overestimated due to the telluric currents not being properly accounted for yet, but they still provide a useful measure of nominal current intensities. The westward edge of the wave-like structure is often brighter than the eastward edge of the omega form (as in the right panel of Fig. 6). Similar more intense trailing flank of the omega form was previously reported by Amm et al. (2005). They concluded that the higher emission intensity indicates a higher number flux of precipitating particles.

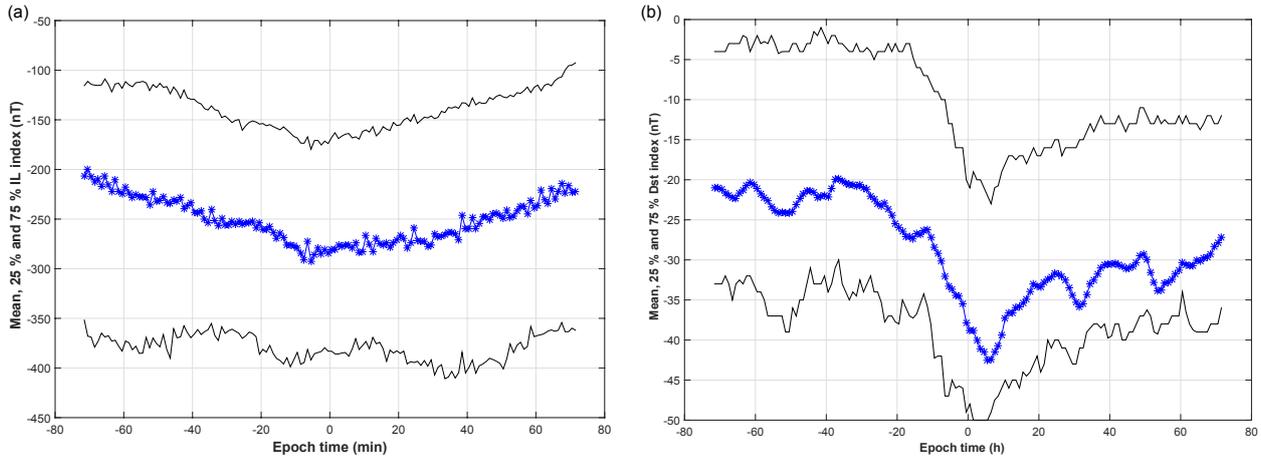

**Figure 5.** Median epoch values of local electrojet index (IL$_{ASC}$ in panel **a**) and Dst index data (**b**) as the blue curve. The epoch time of the IL index is in minutes and the one of Dst index is in hours due to the different temporal resolution of the index data. The zero epoch time is the individual omega peak time for both indices. We have included all omega-like structures in the superposed epoch analysis. Range of variations in index values is described by quartiles (black curves).

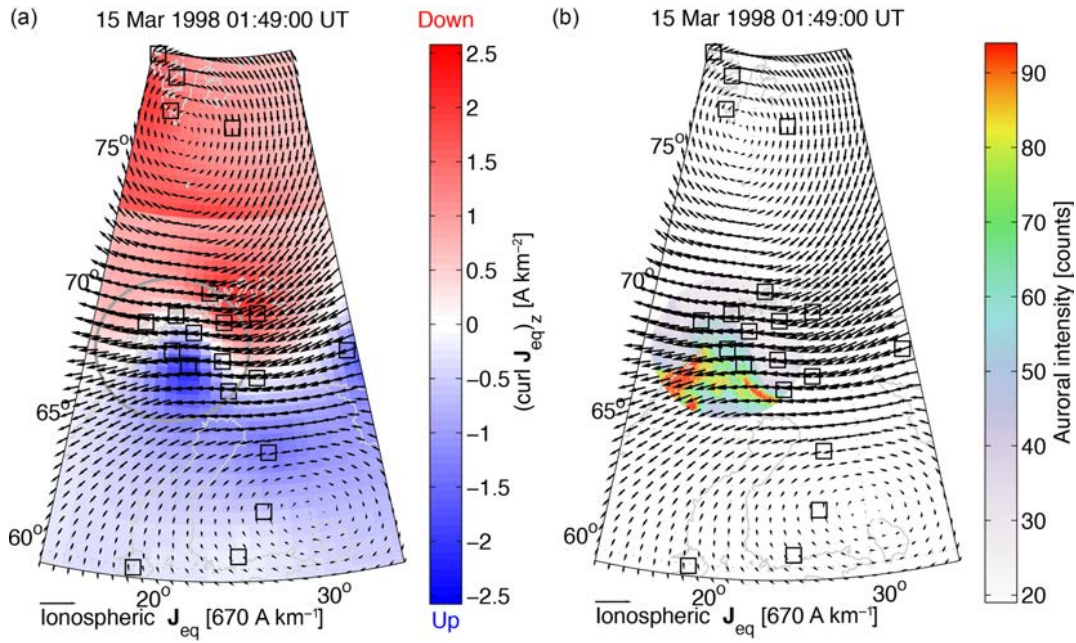

**Figure 6.** Distribution of equivalent currents (vectors in 670 A km$^{-1}$) for an omega event on 15 March 1998 at 01:49 UT. The curl of the equivalent currents gives an estimate of the field-aligned current distribution (**a**) downward (red) and upward (blue) in A km$^{-2}$. An ASC image of an auroral omega structure from ABK camera has been plotted on the equivalent current vector field (**b**) in relative brightness units and false colour (counts).

The omega bands are further related to upward field-aligned current (FAC), which is estimated as the equivalent current vorticity. The boundary between upward and downward FAC undulates in the same manner as the optical emission boundary does. All reference set omegas (259 individual forms) were visually examined and seen to propagate as undulations at the boundary of the equivalent current vorticity. It has been shown by Weygand et al. (2016) that even if the curl of the equivalent current has its limitation as a FAC proxy, the boundary of the vorticity field describes the FAC boundary well. In particular, the brighter trailing edge of the omega form follows the FAC boundary very well. The good correspondence between the diffuse boundary of the optical aurora and the boundary of the equivalent current vorticity is further illustrated by the ASC keogram (false-colour) and the equivalent current keogram along the camera station lon-gitude (Fig. 7). The evolution of the current boundary and the poleward boundary of the emission are in a good agreement until the passage of the last omega structure at about 02:10 UT.

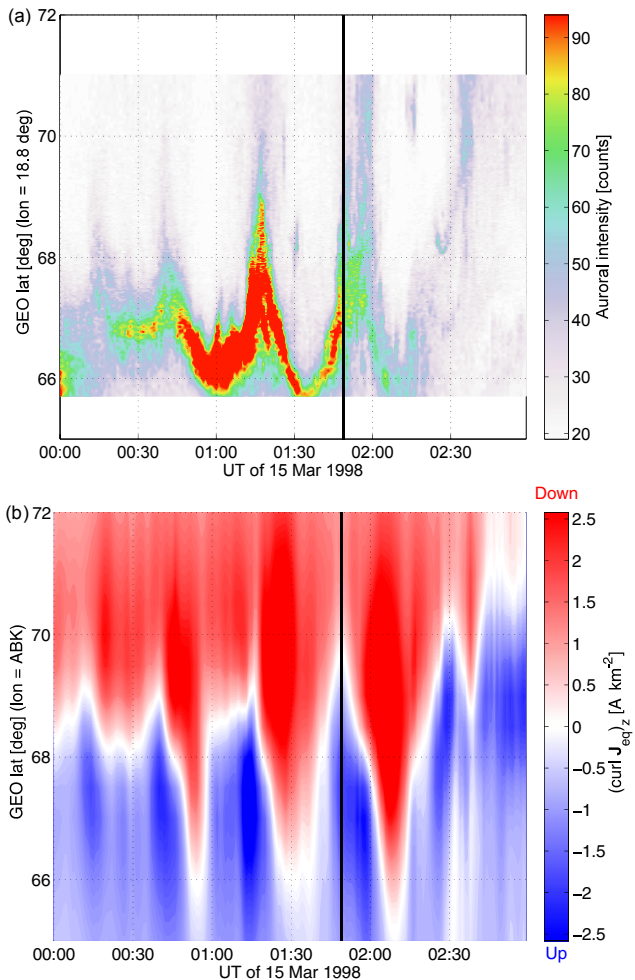

**Figure 7.** Evolution of the auroral brightness **(a)** and the equivalent current vorticity **(b)** as a function of time and geographic latitude for an omega event on 15 March 1998. The vertical line marks the individual omega structure in Fig. 6 at 01:49 UT. Other omega structures took place at 00:12, 00:17 and 00:40 UT. The curl of the equivalent currents gives an estimate of the field-aligned current distribution downward (red) and upward (blue) in $A\,km^{-2}$. ASC data are plotted in relative brightness units (counts) and false colour. Both keograms are plotted along the longitude of the ABK camera station (18.8°).

The selected event on 15 March 1998 consists of three distinct (at 00:17, 00:40 and 01:49 UT) and one less-obvious (at 00:12 UT) individual omega structures detected at ABK station. All omega forms are large. For smaller omegas the corresponding equivalent current evolution is less pronounced. A similar dependence between auroral brightness and ionospheric currents was also observed by Amm et al. (2005). However, the equivalent current estimates by Amm et al. (2005) were an order of magnitude stronger than the ones in our sample omega. Their event took place during an intense geomagnetic storm while our sample event is related to moderate activity with Dst about $-40$ nT and AE about 400–500 nT. FACs strengths of the order of $1\,A\,km^{-2}$ were also calculated by a similar method in the study by Weygand et al. (2015), in which the geomagnetic activity was comparable to that of our sample event.

## 5 Spacecraft observations

Of all the omega band events, we found four periods with near-conjugate spacecraft data in the plasma sheet. These spacecraft observations come from Geotail (Nishida et al., 1997). Below we show spacecraft observations for two of the four events, for which the spacecraft footpoint was closest to the ground-based ASC FoV. In the other two omega events, no signs of earthward fast flows (15 October 2001) were observed, or short-lived fast flows (29 March 2001) were observed after the period of the auroral omega band. Both of these excluded events occurred in less-optimal conjugacy between the ground and space-based measurements.

### 5.1 29 January 1998

An omega band was observed in the ABK and KIL ASC images from 22:21 to 22:32 UT. Sample images of one (and only) individual omega structure from both stations are shown in Fig. 8. The formation of the omega band was not observed during this event but the omega structure appeared to increase in height as it drifted from west to east. During this period the Geotail spacecraft was located at about $(-30.5, -0.4, -2.5)$ $R_E$ GSM and the footpoint mapped with the T96 model to about 2° to the northwest of the ABK and KIL stations, just outside the FoV of ABK at 70.2° Glat, 9.7° Glong. Using the T01 model the footpoint of the Geotail spacecraft maps to 72.4° Glat, 357.8° Glong, which is further to the west and to the north than the T96 footpoint. The omega band did not cross through the T96 or T01 footpoint of the Geotail spacecraft. This is the closest conjugate event on our event list. During this period the IL index reaches a minimum of about $-700$ nT. Figure 9 shows the magnetic field at about 1 min resolution (top panel) and Hot Plasma Analyzer (Frank et al., 1994) observations (panels 2–4) during this period. The most significant features visible are the high-speed earthward flows from about 22:15 UT to about 22:22 UT with a peak speed of about $500\,km\,s^{-1}$ and a slower earthward flow of a little more than $200\,km\,s^{-1}$ at 22:31 UT (second panel). We note that there is one large high-speed flow and only one omega was observed in the ionosphere. This event supports the theory that streamers (high-speed earthward flow in the plasma sheet) would be the mechanism by which omega bands are created. However, the spatial correlation between the measured flow and the omega structure is not one to one, there is a large difference between the T01 and T96 footpoint locations, and no streamers can be found in the auroral images prior to the omega formation.

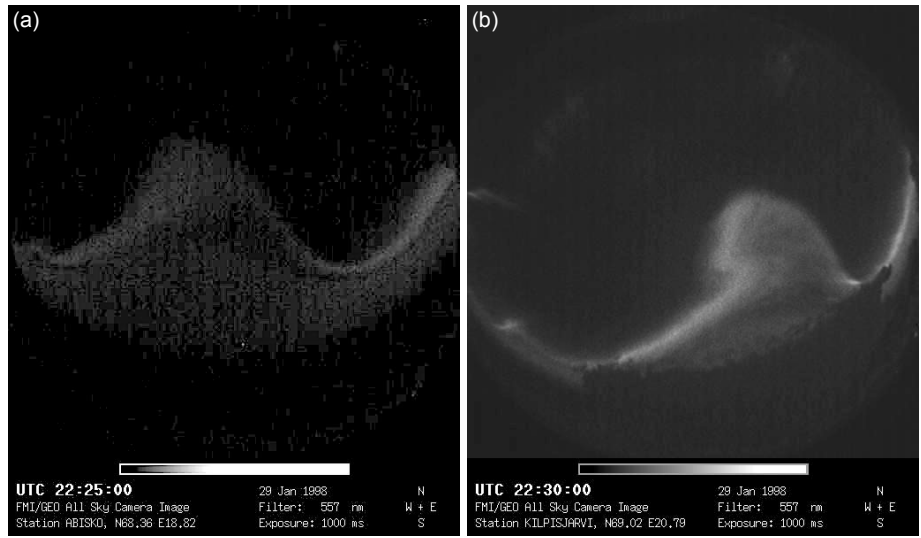

**Figure 8.** Auroral omega forms observed at ABK at 22:25 UT **(a)** and KIL at 22:30 UT **(b)**. The contrast in the images has been enhanced.

## 5.2  4 December 1999

An omega band was observed in the KEV, KIL and MUO all-sky images from 00:12 to 00:19 UT. Sample images of the two individual omega forms of this event are shown in Fig. 10.

The omega band propagated from the west to the east into the FoV of the KIL images then across the KEV and MUO FoV. During the eastward propagation the omega appeared to thicken and increase in height, but we did not see the omega form. During this period Geotail was located at about $(-17.9, -8, -3.2)$ $R_\mathrm{E}$ GSM and the footpoint mapped with the T96 model to about 71.5° Glat, 5.1° Glong, which is 15° west of the KIL station and outside the ASC FoVs. The T01 model maps the footpoint to 70.0° Glat, 353° Glong. The footpoint of the spacecraft did not cross through the omega band for either magnetic field line model. Figure 11 shows the magnetic field (top panel) and Hot Plasma Analyzer observations (panels 2–4) during this period. The most significant features visible are the high-speed tailward flows from about 00:07 UT to about 00:23 UT with a maximum speed of about $-500\,\mathrm{km\,s^{-1}}$ (second panel), and from 00:12 to 00:19 UT there are $V_Y$ flows greater than $250\,\mathrm{km\,s^{-1}}$ (third panel). The $V_X$ flow measurements suggest that reconnection occurred somewhere between the spacecraft and the Earth and there may be equally strong flows earthward of Geotail. Note that between about 00:10 UT and about 00:20 UT the $B_X$ GSM component is about $-20\,\mathrm{nT}$, which could indicate that Geotail is in the southern lobe region making it more difficult to interpret the observed flows. The density measurements varying between 0.20 and $0.35\,\mathrm{cm^{-3}}$, however, do not support that observation, since typical tail lobe values remain below $0.1\,\mathrm{cm^{-3}}$ (Svenes et al., 2008).

## 6  Discussion

During the search through auroral omega band images a large amount of optical data has been viewed. In the auroral images, we see the diffuse aurora boundary undulations drifting into the camera FoV and either drifting eastward or overturning and breaking within the camera FoV. Our impression is that there are plenty of eastward-propagating boundary waves which are very similar to omega-like structures, as defined here, but do not fulfil the strict criteria of omega bands. These omega-like features may manifest a structural pre- or post-omega state and the same physics as the more distinct omega forms. Even if only one aurora boundary undulation has been accepted as an omega during a certain event, a longer series of diffuse auroral boundary waves is typically observed. Boundary undulations both before and after the omega band structures may appear, suggesting that the same physical mechanism is responsible for them all, but what we call an omega or an auroral omega band is only a special case, or maybe a maximum-amplitude wave form in the period of the boundary wave instability. The name may simply be a misnomer due to human expert classification. Thus, an automatic recognition of omega-like features in the future, with somewhat looser definition than in this study, may help in resolving the morphological evolution of auroral omega forms and in acquiring a statistically significant set of conjugate events to give a solid explanation for the physical formation mechanism.

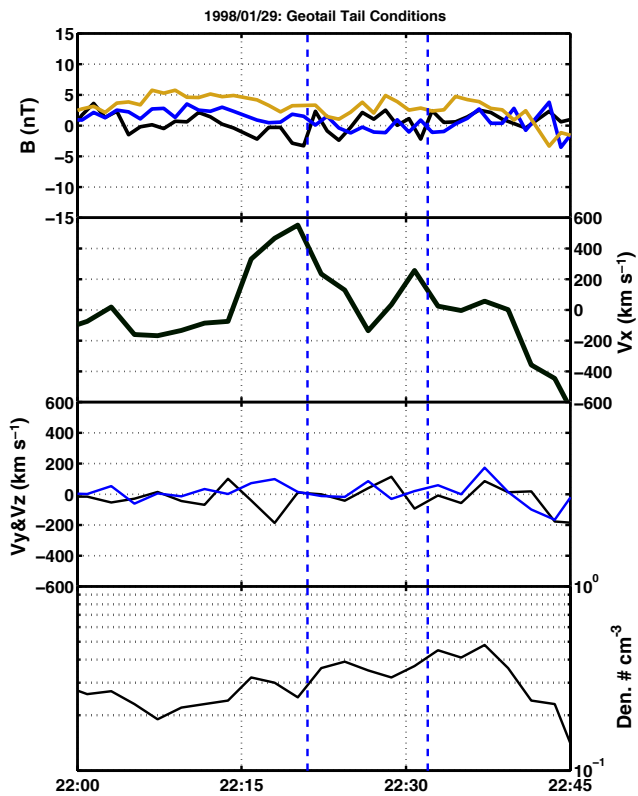

**Figure 9.** Geotail magnetic field and plasma data for 29 January 1998. The top panel shows the magnetic field in GSM coordinates. The black curve is the $B_X$ component, the orange curve is the $B_Y$ component, and the blue curve is the $B_Z$ component. The second panel shows the $V_X$ component of the flow and the third panel displays the $V_Y$ (black curve) and $V_Z$ (blue curve) components. The bottom panel displays the particle density. The dashed vertical lines mark the start and end of the omega band observations at ABK and KIL.

Median Dst values during omega events do not indicate intense storm-time activity but are much too negative for quiet time or average of any substorm phase value in the long-term statistics (Partamies et al., 2013). Similarly, the IL index is lower during the omega observations than during typical substorms in the Lapland region. This finding indicates that a certain activity level beyond that of an average substorm is required for the formation of omegas. An average peak height of auroral emission within the omega forms was found to be about 118 km without significant change during its lifetime. This corresponds to precipitating electron energies of a few keV (Turunen et al., 2009) and suggests that no significant precipitation energy change is involved in the auroral omega evolution. The peak emission-height-based energy estimate agrees with previously estimated precipitation energies from Polar UVI data (Amm et al., 2005) and more recently Cluster data (Wild et al., 2011). Soft particle precipitation and E region peak emission height during omega events may give rise to the observed high electrojet activity. Un-like some other diffuse aurora precipitation, such as pulsating patches (Lessard, 2012; Partamies et al., 2017), omega bands are clearly not related to particularly hard precipitation. However, as pointed out by Weygand et al. (2015) pulsating aurora was seen simultaneously with the omega band activity further east and further equatorward. The events analysed in this study also included omega evolution at the poleward edge while pulsating aurora took place at the equatorward edge of the same diffuse aurora band, if not simultaneously at least within 2 h of omega appearance. This was true for every analysed event where the equatorward boundary of the diffuse aurora was visible in the common FoV of the cameras. More detailed studies are required to investigate the relationship between omega-related instabilities and the acceleration of particles to higher energies of pulsating aurora closer to the Earth.

The majority of the omega forms in this study were observed at 02:00–03:00 MLT and within 1.5 h from the preceding substorm onset in the same local time sector. This suggests that the substorm onsets preceding our omega observations took place mainly at 00:30–01:30 MLT, which is later than the average globally observed substorm onset time of 23:00 MLT (Frey et al., 2004). The previous substorm study of event in the Fennoscandian sector in 1997 and 1999 by Tanskanen et al. (2002) showed that most of the substorms took place within the magnetic latitude range of about 66–67°, which is a couple of degrees higher than the magnetic latitude of SOD station (∼ 64°) where most of our omega bands were observed. Their study also showed an increasing substorm intensity as a function of decreasing latitude. Thus, the latitude difference between omega observations and average substorm onset location agrees with the observation that omega events are associated with more intense than average substorms. Whether the substorm intensity also relates to the later substorm onset time in MLT could be studied in the future.

Structural evolution during the omega bands as described by the arciness index suggests that the passage of an omega band leads to complex auroral structures for at least 20 min after the peak omega time. In only 25 % of omega events, auroral arcs were observed 5 min before. Thus, a great majority of the omega forms develop from a more complex auroral display of substorm expansion and recovery phases. In the substorm expansion phases the omega passage does not change the average auroral complexity of the phase but divides it into arc-like and more complex aurora.

The boundary undulation at the southern part of the auroral oval and westward electrojet region places the magnetospheric counterpart of the omega structures close to the Earth. The transition region between dipolar and tail-like magnetic field configuration is a good candidate for a boundary region where fast earthward flows from the tail reconnection site can launch wave-like instabilities, which would then propagate eastward with the convection return flow. This is in agreement with previous studies (e.g. Weygand et al., 2015)

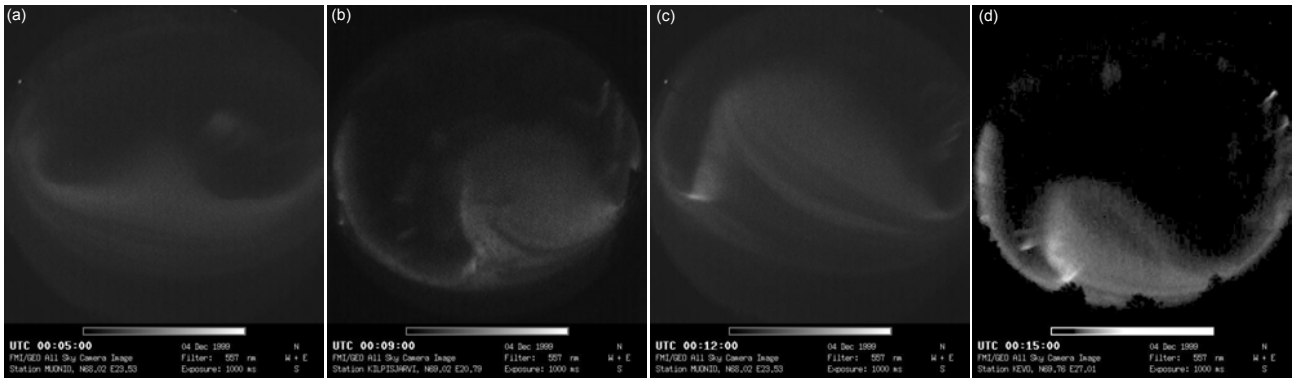

**Figure 10.** Auroral omega forms observed at MUO at 00:05 UT, KIL at 00:09 UT, MUO at 00:12 UT and KEV at 00:15 UT. The first image at 00:05 UT shows the first observed omega form and the last three images capture the second individual omega structure at different stations. The contrast in all images has been enhanced.

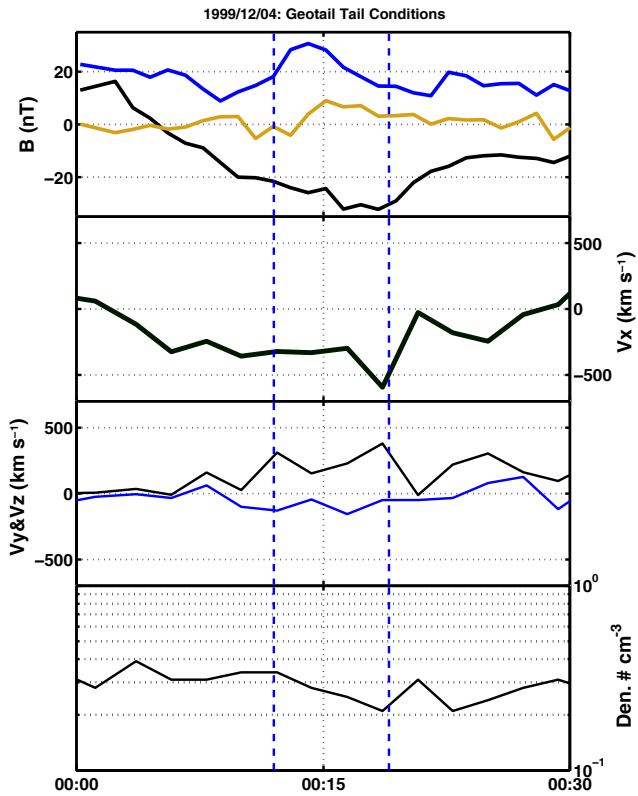

**Figure 11.** This figure has the same format as Fig. 9.

which mapped ionospheric omega structures to just beyond geostationary orbit in the tail.

As suggested by the occurrence of magnetic Ps6 pulsations (Connors et al., 2003), omega bands also occur during the substorm expansion phases and not only during substorm recovery phases. Henderson (2016) concluded that omega bands typically appear in late expansion and early recovery phases, which agrees well with the results in this study. Streamers prior to omegas were not observed in our data, which is interpreted as a limited common FoV of the Lapland ASC stations rather than a non-existent relationship between streamers and omegas. Both of these auroral features are rather short-lived and localized. Reliable and systematic observations of streamers and omegas and their relative appearance would require high-resolution optical observations over much larger spatial area.

Omega band distribution in expansion and recovery phases (Fig. 4) is similar to the magnetotail high-speed distributions for expansion and recovery phases reported by Juusola et al. (2011) for speeds of 400–700 km s$^{-1}$ (their Fig. 4). For these high-speed earthward flows the occurrence rate is steady throughout the expansion phase, it peaks at the beginning of the recovery phase and decays towards the end of the recovery phase. The similarity between the omega and high-speed earthward flow occurrences supports the idea that the omega band formation requires enhanced plasma sheet flows. Earlier studies by, for example, Henderson et al. (2012), Weygand et al. (2015) and Henderson (2016) show the north–south-aligned auroral streamers develop to torches and further to omega bands after reaching the equatorward part of the auroral oval, which reflects an increase in plasma sheet earthward flows before fully formed omegas can be observed in the ionosphere. This may require strong enough plasma flow to reach the more dipolar field at the earthward edge of the plasma sheet. Magnetospheric flow speeds during the best conjugate event of this study (29 January 1998) are high indeed. The uncertainty of the current analysis is that the observed conjugate spacecraft are not spatially crossing the optically observed omega structures. Thus, no solid conclusions can be drawn without a larger spacecraft–ground-based conjugate study. Although the omega bands are generally large in the ASC FoV, the well-defined structures are rather rare and transient, which makes it unlikely for a spacecraft to pass right through. Furthermore, a full morphological evolution of omega bands is hard to capture due to the large-scale size and fast propagation of these structures, but it would be es-sential in understanding any near-conjugate space-borne observations.

## 7 Conclusions

We have used a combination of automated and visual search methods to detect 438 auroral omega structures in MIRACLE all-sky camera data from five identical Lapland stations in 1996–2007. This is the largest statistical omega study to date which includes analysis of not only the occurrence time but also typical lifetime and relation to geomagnetic activity.

We conclude that both local and global magnetic indices indicate substorm activity which is stronger than average but still not at magnetic storm-time levels. Based on equivalent current maps, the omega bands appear at the boundary of region 1 and 2 currents and within westward electrojets which are more intense than those observed during average substorms. Average peak emission height of 118 km in the E region supports the observations of enhanced electrojet activity. Furthermore, steady height values suggest a constant average precipitation energy throughout the evolution of the omega events.

The local time distribution of omega bands peaks at the morning hours (at 02:00–03:00 MLT), and the annual occurrence rate is highest in the declining phase of the solar cycle (in 2002–2004). The omega occurrence in substorm phases shows a steady distribution in expansion phase, peak at the beginning of the recovery phase and a decline towards the end of the recovery phase. This behaviour is similar to earlier reported occurrence rate of high-speed plasma sheet flows in substorm expansion and recovery phases, and supports the idea that high-speed earthward flows in the tail are associated with the formation of auroral omega bands. North–south-aligned auroral streamers were typically not observed prior to the omega bands in our data set. This, however, does not mean that no streamers occurred, but rather that the possible streamer activity did not take place within the MIRACLE ASC FoVs in the same region where the omegas occurred. Out of four tail conjugate events only one was associated with spacecraft measurements of fast earthward flows. None of our spacecraft–ground-based conjugate events mapped exactly onto one another.

*Data availability.* The Dst index data were downloaded from Kyoto World Data Center (2017) at http://wdc.kugi.kyoto-u.ac.jp. MIRACLE ASC quicklook data (FMI, 2017) are available at http://space.fmi.fi/MIRACLE/ASC/index.html and full-resolution images upon request from Kirsti Kauristie at FMI (kirsti.kauristie@fmi.fi).

*Competing interests.* The authors declare that they have no conflict of interest.

*Acknowledgements.* The authors thank Sanna Mäkinen, Jyrki Mattanen, Anneli Ketola, Tero Raita and Carl-Fredrik Enell for careful maintenance of the camera network and data flow, and Mikko Syrjäsuo for implementing the random projection method for ASC data. The collaboration work of Noora Partamies and James M. Weygand was supported by the Norwegian Research Council grant 223252 and UNIS.

The topical editor, Christopher Owen, thanks Michael G. Henderson and one anonymous referee for help in evaluating this paper.